\documentclass[12pt]{article}
\usepackage{amsfonts}

\usepackage{graphicx}
\usepackage{amsmath}
\usepackage{float}

\textwidth=6.5in \hoffset=-.75in \voffset=-1in \numberwithin{equation}{section}
\numberwithin{figure}{section}

\begin{document}

\begin{titlepage}
\bigskip \begin{flushright}
\end{flushright}
\vspace{1cm}
\begin{center}
{\Large \bf {Supergravity Solutions Without Tri-holomorphic $U(1)$ Isometries}}\\
\end{center}
\vspace{1cm}
\begin{center}
 A. M.
Ghezelbash{ \footnote{ E-Mail: masoud.ghezelbash@usask.ca}}
\\
Department of Physics and Engineering Physics, \\ University of Saskatchewan, 
Saskatoon, Saskatchewan S7N 5E2, Canada\\
\vspace{1cm}
\end{center}
\begin{abstract}

We investigate the construction of five-dimensional supergravity solutions that don't have 
any tri-holomorphic $U(1)$ isometries.
We construct a class of solutions that in various limits of parameters reduces to many of previously constructed
five-dimensional supergravity solutions based on both hyper-K\"{a}hler base spaces
that can be put into a Gibbons-Hawking form and hyper-K\"{a}hler base spaces that
can't be put into a Gibbons-Hawking form.
We find a new solution which is over triaxial Bianchi type IX
Einstein-hyperk\"{a}hler base space with no tri-holomorphic $U(1)$ symmetry.
One special case of this solution corresponds to
five-dimensional solution based on Eguchi-Hanson 
type II geometry. 
\end{abstract}
\bigskip
\hspace*{1cm} PACS: 04.65.+e, 04.50.-h, 11.30.-j
\end{titlepage}\onecolumn

\bigskip

\section{Introduction}

It is believed that in the strong coupling limit, many horizonless
three-charge brane configurations undergo a geometric transition and become
smooth horizonless geometries with black hole or black ring charges \cite{B1}%
. These charges come completely from fluxes wrapping on non-trivial cycles.
The three-charge black hole (ring) systems are dual to the states of
corresponding conformal field theories: in favor of the idea that non-fundamental-black hole
(ring) systems effectively arise as a result of many horizonless
configurations \cite{B2,Ma1}. The simplest eleven-dimensional supergravity
metrics (the low-energy limit of M-theory \cite{gr2}) with three-charge
geometries have the form \cite{B3}%
\begin{eqnarray}
ds_{11}^{2} &=&-(Z_{1}Z_{2}Z_{3})^{-2/3}(dt+\omega
)^{2}+(Z_{1}Z_{2}Z_{3})^{1/3}ds_{4}^{2}  \label{11d} \\
&+&(\frac{Z_{2}Z_{3}}{Z_{1}^{2}})^{1/3}(dx_{5}^{2}+dx_{6}^{2})+(\frac{%
Z_{3}Z_{1}}{Z_{2}^{2}})^{1/3}(dx_{7}^{2}+dx_{8}^{2})+(\frac{Z_{1}Z_{2}}{%
Z_{3}^{2}})^{1/3}(dx_{9}^{2}+dx_{10}^{2}),  \notag
\end{eqnarray}%
where $x_{5},...,x_{10}$ are coordinates of compactification six-torus and $%
ds_{4}^{2}$ is any hyper-K\"{a}hler metric (which is equivalent to a metric
with self-dual curvature in four dimensions). To preserve supersymmetry, the
base metric $ds_{4}^{2}$ should be hyper-K\"{a}hler \cite{G1}. The
five-dimensional space-time sub-metric of (\ref{11d}) together with Maxwell
field make the bosonic sector of five-dimensional minimal supergravity. In
five-dimensions, unlike the four dimensions that the only horizon topology
is 2-sphere, we can have different more interesting horizon topologies such
as black holes with horizon topology of 3-sphere \cite{Myers}, black rings
with horizon topology of 2-sphere $\times $ circle \cite{Em1,Em2}, black
saturn: a spherical black hole\ surrounded by a black ring \cite{El1}, black
lens which the horizon geometry is a Lens space $L(p,q)$ \cite{Ch1}. All
allowed horizon topologies have been classified in \cite{Ca1,He1,Ga1}.
Recently, it was shown how a uniqueness theorem might be proved for black
holes in five dimensions \cite{Ho1,Ho2}. It was shown stationary,
asymptotically flat vacuum black holes with two commuting axial symmetries
are uniquely determined by their mass, angular momentum and rod structure.
Specifically, the rod structure \cite{Ha1} determines the topology of
horizon in five dimensions.

In \cite{B4}, the authors have used hyper-K\"{a}hler Atiyah-Hitchin base
space and its ambi-polar generalizations to construct five-dimensional,
three-charge supergravity solutions that only have a rotational $U(1)$
isometry.\ The complete solutions are regular around the critical surface of
ambi-polar base space. These solutions are very remarkable because
(ambi-polar) hyper-K\"{a}hler Atiyah-Hitchin geometries (unlike ambi-polar
Gibbons-Hawking geometries) don't have any tri-holomorphic $U(1)$ isometry
(tri-holomorphic $U(1)$ isometry means the $U(1)$ preserves all three complex structures of the hyper-K%
\"{a}hler geometry). Hence they could be used to study the interesting
physical processes such as, merger
of two Breckenridge-Myers-Peet-Vafa black holes \cite{Br1} or the geometric
transition of a three-charge supertube of arbitrary shape; that don't respect any tri-holomophic $U(1)$ symmetry. In \cite{KH}, the authors also have used the Atiyah-Hitchin metric and constructed a solution to five-dimensional minimal
supergravity.

The Atiyah-Hitchin metric is a special case of Bianchi type IX Einstein-K\"{a}hler metrics
with
generic non-tri-holomorphic $U(1)$ isometries. It's interesting to note that
in some special limits, 
Bianchi type IX space reduces to
Taub-NUT and Eguchi-Hanson spaces (the latter geometry will be referred as Eguchi-Hanson type II in this
paper).

The Bianchi type IX space has been used in construction of M2 and M5 brane
solutions which are realization of supergravity solutions for localized IIA
D2/D6(2), NS5/D6(5) and IIB NS5/D5(4) intersecting brane systems \cite{Gh1}.
By lifting a D6 (D5 or D4)-brane to four-dimensional hyper-K\"{a}hler
Bianchi type IX geometry embedded in M-theory, these solutions were
constructed by placing M2- and M5-branes in the Bianchi type IX background
geometry. The special feature of this constructions\ is that the solutions
are not restricted to be in the near core region of the D6 (D5 or
D4)-brane.\ \ Moreover, all of the different solutions preserve 1/4 of the
supersymmetry as a result of the self-duality of the Bianchi type IX metric.
All previously known M2 and M5 solutions \cite{hashi,CGMM2,CGMM22,ATM2} are
special cases of the solutions presented in \cite{Gh1}.

The Bianchi type IX spaces were used, recently, for construction cohomogeneity
two metrics of $G_{2}$ holonomy which are foliated by twistor spaces \cite%
{CV}. The twistor spaces are two-sphere bundles over Bianchi type IX
Einstein metrics with self-dual Weyl tensor.

In this paper, we use self-dual Bianchi type IX space as the base space to
construct some new five-dimensional supergravity solutions with generic non-tri-holomorphic $U(1)$ isometries. 
We would like to stress 
that, in general, constructing solutions with non-tri-holomorphic $U(1)$ isometries is a rather complicated, 
tedious and challenging task. 
To our knowledge, for classical black holes and black rings, only two solutions exist \cite{B11,B22}. The outline of this paper is as follows. In section \ref{sec:5Dreview},
we give a brief review of five-dimensional supergravity and equations of motion. In section \ref{sec:BIX}, we present Bianchi type IX space and show how the space reduces to different well-known spaces that were used previously for constructing five-dimensional supergravity solutions. We consider in detail
the case of triaxial Bianchi type IX space that could be considered for construction of a new class of supergravity solutions. In section \ref{sec:sc}
we consider a class of supergravity solutions over triaxial Bianchi type IX space and present the analytic expressions for the solutions near the center of space-time and also at infinity. We also provide the results of numerical integration of equations of motion and discuss the behaviors of our solutions.

\section{Five-dimensional Minimal Supergravity}

\label{sec:5Dreview}

The bosonic part of five-dimensional minimal supergravity is
Einstein-Maxwell theory with a Chern-Simon term and is given by the action %
\cite{Gau1}
\begin{equation}
S=\frac{-1}{4\pi G}\int (\frac{1}{4}R\ast 1+\frac{1}{2}F\wedge \ast F+\frac{2%
}{3\sqrt{3}}F\wedge F\wedge A).  \label{action}
\end{equation}%
The equations of motion are%
\begin{equation}
R_{\mu \nu }+2F_{\mu \lambda }F_{\nu }^{\text{ \ }\lambda }-\frac{1}{3}%
g_{\mu \nu }F^{2}=0,  \label{eq1}
\end{equation}%
\begin{equation}
d\ast F+\frac{2}{\sqrt{3}}F\wedge F=0.  \label{eq2}
\end{equation}

A bosonic solution is supersymmetric if it admits a super-covariantly
constant, symplectic Majorana Killing spinor $\varepsilon ^{\mu }$ obeying%
\begin{equation}
D_{\mu }\varepsilon ^{a}+\frac{1}{4\sqrt{3}}(\gamma _{\mu }^{\text{ \ \ }\nu
\lambda }-4\delta _{\mu }^{\nu }\gamma ^{\lambda })F_{\nu \lambda
}\varepsilon ^{a}=0.  \label{killing}
\end{equation}%
From a single commuting $\varepsilon ^{a},$ a scalar $f,$ a 1-form $V,$ and
three 2-forms $\Phi ^{ab}\equiv \Phi ^{(ab)}$ could be constructed \cite%
{Gau1}; given by%
\begin{equation}
f\varepsilon ^{ab}=\overline{\varepsilon }^{a}\varepsilon ^{b},  \label{f}
\end{equation}%
\begin{equation}
V_{\mu }\varepsilon ^{ab}=\overline{\varepsilon }^{a}\gamma _{\mu
}\varepsilon ^{b},  \label{v}
\end{equation}%
\begin{equation}
\Phi _{\mu \nu }^{ab}=\overline{\varepsilon }^{a}\gamma _{\mu \nu
}\varepsilon ^{b}.  \label{phi}
\end{equation}

The solutions could be classified depending on the Killing vector $V_{\mu }$
to be timelike or null. If we consider the case in which $f$ is not zero and 
$V=\frac{\partial }{\partial t}$ is a timelike Killing vector field, then
the metric\ can be written as
\begin{equation}
ds^{2}=-f^{2}(dt+\omega )^{2}+\frac{1}{f}ds_{B}^{2},  \label{5dm}
\end{equation}%
where $ds_{B}^{2}=h_{mn}dx^{m}dx^{n}$ is the metric of the four-dimensional
hyper-K\"{a}hler base space $B$ \cite{Gau1}. We note that the metric $\frac{1%
}{f}ds_{B}^{2}$ is obtained by projecting the full five-dimensional metric $%
ds^{2}$ perpendicular to the orbits of Killing vector field $V.$

If we define 
\begin{equation}
{\bold e}^{0}=f(dt+\omega ),  \label{e0}
\end{equation}%
then ${\bold e}^{0}\wedge {\bold \eta} $ defines a positive orientation for the
five-dimensional metric (\ref{5dm}), where ${\bold \eta} $ is a positive orientation
on the base space $B.$ The two form $d\omega $ only has components tangent
to the base space $B$ and so it can be split into self-dual and
anti-self-dual parts with respect to the metric of base space%
\begin{equation}
d\omega =\frac{G^{+}}{f}+\frac{G^{-}}{f}.  \label{domega}
\end{equation}%
Taking the differentials of $f$ and $V$ and then the exterior derivatives of
obtained equations and using the fact that $V$ is a Killing vector, leads to
the following result for the two form $F$%
\begin{equation}
F=\frac{\sqrt{3}}{2}(-\frac{1}{f^{2}}V\wedge df+\frac{1}{3}G^{+}+G^{-}),
\label{Feq}
\end{equation}%
or%
\begin{equation}
F=\frac{\sqrt{3}}{2}de^{0}-\frac{1}{\sqrt{3}}G^{+},  \label{Feq2}
\end{equation}%
where $G^{+}$ is given by%
\begin{equation}
G^{+}=\frac{f}{2}(d\omega +\ast _{B}d\omega ).  \label{Gplus}
\end{equation}

From the Bianchi identity and equation of motion (\ref{eq2}), we get the following
equations%
\begin{equation}
dG^{+}=0,  \label{dg}
\end{equation}%
\begin{equation}
\bigtriangledown ^2 \frac{1}{f}=\frac{4}{9}(G^{+})^{2}=\frac{2}{9}%
(G^{+})_{mn}(G^{+})^{mn},  \label{eqforF}
\end{equation}%
where $\bigtriangledown ^2 $ is the Laplace operator on base space $B$.

\section{The (Triaxial) Bianchi Type IX Space}

\label{sec:BIX}

The Bianchi type IX metric $ds_{B.\text{ }IX}^{2}$ is locally given by the
following metric with an $SU(2)$ or $SO(3)$ isometry group \cite{GM} 
\begin{equation}
ds_{B.\text{ }IX}^{2}=e^{2\{A(\zeta )+B(\zeta )+C(\zeta )\}}d\zeta
^{2}+e^{2A(\zeta )}\sigma _{1}^{2}+e^{2B(\zeta )}\sigma _{2}^{2}+e^{2C(\zeta
)}\sigma _{3}^{2},  \label{BIXG}
\end{equation}%
where $\sigma _{i}$'s are Maurer-Cartan one-forms (see Appendix). The metric (\ref{BIXG}) satisfies Einstein
equations provided%
\begin{eqnarray}
\frac{d^{2}A}{d\zeta ^{2}}&=&\frac{1}{2}\{e^{4A}-(e^{2B}-e^{2C})^{2}\}, \label{App}\\ 
\frac{d^{2}B}{d\zeta ^{2}}&=&\frac{1}{2}\{e^{4B}-(e^{2C}-e^{2A})^{2}\}, \label{Bpp}\\ 
\frac{d^{2}C}{d\zeta ^{2}}&=&\frac{1}{2}\{e^{4C}-(e^{2A}-e^{2B})^{2}\},\label{Cpp}%
\end{eqnarray}%
\newline
and%
\begin{equation}
\frac{dA}{d\zeta }\frac{dB}{d\zeta }+\frac{dB}{d\zeta }\frac{dC}{d\zeta }+%
\frac{dC}{d\zeta }\frac{dA}{d\zeta }=\frac{1}{2}%
\{e^{2(A+B)}+e^{2(B+C)}+e^{2(C+A)}\}-\frac{1}{4}\{e^{4A}+e^{4B}+e^{4C}\}.
\label{Ein2}
\end{equation}%
Moreover self-duality of the curvature implies 
\begin{eqnarray}
\frac{dA}{d\zeta }&=&\frac{1}{2}\{e^{2B}+e^{2C}-e^{2A}\}-\alpha _{1}e^{B+C},\label{Ap} \\ 
\frac{dB}{d\zeta }&=&\frac{1}{2}\{e^{2C}+e^{2A}-e^{2B}\}-\alpha _{2}e^{A+C}, \label{Bp}\\ 
\frac{dC}{d\zeta }&=&\frac{1}{2}\{e^{2A}+e^{2B}-e^{2C}\}-\alpha _{3}e^{A+B},\label{Cp}%
\end{eqnarray}%
where three constant numbers $\alpha _{i},i=1,2,3$ satisfy%
\begin{equation}
\alpha _{i}\alpha _{j}=\varepsilon _{ijk}\alpha _{k}.  \label{alphas}
\end{equation}%
We note that equations (\ref{Ap}),(\ref{Bp}) and (\ref{Cp}) are integrals of equations 
(\ref{App}), (\ref{Bpp}), (\ref{Cpp}) and (\ref{Ein2}%
). The possible solutions of equation (\ref{alphas}) are\newline

\hspace{-0.75cm}
I) $(\alpha _{1},\alpha _{2},\alpha _{3})=(1,1,1)$\newline
II) $(\alpha _{1},\alpha _{2},\alpha _{3})=(1,-1,-1)$\newline
III) $(\alpha _{1},\alpha _{2},\alpha _{3})=(-1,1,-1)$\newline
IV) $(\alpha _{1},\alpha _{2},\alpha _{3})=(-1,-1,1)$\newline
V) $(\alpha _{1},\alpha _{2},\alpha _{3})=(0,0,0)$\newline

Here we consider all five cases:\newline

{\underline {Case I}}:\newline

Choosing $(\alpha _{1},\alpha _{2},\alpha _{3})=(1,1,1)$ in equations (\ref{Ap}),
(\ref{Bp}) and (\ref{Cp}) yields the Atiyah-Hitchin metric \cite{Hana} in the form of (\ref{BIXG})
with%
\begin{eqnarray}
e^{2A(\zeta )}&=&\frac{2}{\pi }\frac{\vartheta _{2}\vartheta _{3}^{^{\prime
}}\vartheta _{4}^{^{\prime }}}{\vartheta _{2}^{^{\prime }}\vartheta
_{3}\vartheta _{4}},\label{e2A} \\ 
e^{2B(\zeta )}&=&\frac{2}{\pi }\frac{\vartheta _{2}^{^{\prime }}\vartheta
_{3}\vartheta _{4}^{^{\prime }}}{\vartheta _{2}\vartheta _{3}^{^{\prime
}}\vartheta _{4}},\label{e2B} \\ 
e^{2C(\zeta )}&=&\frac{2}{\pi }\frac{\vartheta _{2}^{^{\prime }}\vartheta
_{3}^{^{\prime }}\vartheta _{4}}{\vartheta _{2}\vartheta _{3}\vartheta
_{4}^{^{\prime }}},\label{e2C}%
\end{eqnarray}%
where the $\vartheta $'s are Jacobi Theta functions with complex modulus $%
i\zeta $. The Jacobi Theta functions are given explicitly in the Appendix.
Since the Atiyah-Hitchin base space and its ambi-polar generalizations have
been considered in explicit construction of the most general
five-dimensional supersymmetric solutions \cite{B4}, so we don't study
them in this article.\newline

{\underline {Cases II, III and IV}}:\newline

All these three cases are not distinct from case I, since they could be obtained
by substitutions, $e^A \rightarrow -e^A$, $e^B \rightarrow -e^B$ and $e^C \rightarrow -e^C$
respectively.\newline

{\underline {Case V}}:\newline

By choosing $(\alpha _{1},\alpha _{2},\alpha _{3})=(0,0,0)$ the differential
equations (\ref{App}), $\cdots$, (\ref{Cp}) can be solved exactly. We find the solutions
\begin{eqnarray}
A(\zeta)&=&\frac{1}{2}\ln \big ( {c^{2}
\frac{\mathfrak{cn}(c^2\zeta ,k^2)\mathfrak{dn}(c^2\zeta ,k^2)}{\mathfrak{sn}(-c^2\zeta ,k^2)}}
\big ), \label{A1} \\
B(\zeta)&=&\frac{1}{2}\ln \big ( {c^{2}
\frac{\mathfrak{cn}(c^2\zeta ,k^2)}{\mathfrak{dn}(c^2\zeta ,k^2)\mathfrak{sn}(-c^2\zeta ,k^2)}}
\big ),\label{A2} \\
C(\zeta)&=&\frac{1}{2}\ln \big ( {c^2
\frac{\mathfrak{dn}(c^2\zeta ,k^2)}{\mathfrak{cn}(c^2\zeta ,k^2)\mathfrak{sn}(-c^2\zeta ,k^2)}}
\big ),\label{A3} 
\end{eqnarray}
where $\mathfrak{sn}(z,k)$, $\mathfrak{cn}(z,k)$ and $\mathfrak{dn}(z,k)$ are the standard Jacobi elliptic $SN$, $CN$ and $DN$ functions. We review Jacobi Theta functions as well as Jacobi elliptic functions lore in the Appendix. 
We change the 
coordinate $\zeta $ in the metric (\ref{BIXG})
to the coordinate $r$ by%
\begin{equation}
r=\frac{2c}{\sqrt{\mathfrak{sn}(c^2\zeta ,k^2)}},  \label{rzeta}
\end{equation}%
hence, we find the metric in the form (that we call it as triaxial Bianchi type IX space) \cite{Gh1}%
\begin{equation}
ds_{tri.\text{ }B.\text{ }IX}^{2}=\frac{dr^{2}}{\sqrt{F(r)}}+\frac{r^{2}}{4}%
\sqrt{F(r)}\left( \frac{\sigma _{1}^{2}}{1-\frac{a_{1}^{4}}{r^{4}}}+\frac{%
\sigma _{2}^{2}}{1-\frac{a_{2}^{4}}{r^{4}}}+\frac{\sigma _{3}^{2}}{1-\frac{%
a_{3}^{4}}{r^{4}}}\right),   \label{BIX}
\end{equation}%
where
\begin{equation}
F(r)=\prod_{i=1}^{3}(1-\frac{a_{i}^{4}}{r^{4}}),  \label{FF}
\end{equation}%
and $a_{1},a_{2}$ and $a_{3}$ are three parameters that with no loss of
generality, we choose them such that $0=a_{1}\leq a_{2}=2kc\leq $ $a_{3}=2c$%
. We note that the coordinate $r$ must be greater or equal to $a_3$. \ Here $0\leq k\leq 1$ is the square root of 
modulus of different types of Jacobi
elliptic functions and $c>0.$ If $k>1,$ all we need is just to interchange
the $2$ and $3$ directions. The metric function (\ref{FF}) is positive definite for $r\geq a_3=2c$, and 
the change of coordinate (\ref{rzeta}) guarantees this requirement. For simplicity, we choose coordinate $\zeta$ in
the range $[0,\alpha_{(c)(k)(1)}]$ where $\alpha_{(c)(k)(m)}$ is the m-th positive root of 
${\mathfrak{sn}(c^2\zeta ,k^2)}$. Any other range of the form $[\alpha_{(c)(k)(2n)},\alpha_{(c)(k)(2n+1)}]$ with 
$n=1,2,3,\cdots$
or
$[-\alpha_{(c)(k)(2n)},-\alpha_{(c)(k)(2n-1)}]$ can be chosen equivalently.

For the special value of $k=0$, where the smaller two $a$'s coincide, we
find the metric (\ref{BIX}) reduces to the following metric%
\begin{equation}
ds_{EHI}^{2}=\frac{dr^{2}}{h(r)}+\frac{r^{2}}{4}h(r)\{d\theta ^{2}+\sin
^{2}\theta d\phi ^{2}\}+\frac{r^{2}}{4h(r)}(d\psi +\cos \theta d\phi )^{2},
\label{EHI}
\end{equation}%
which is the Eguchi-Hanson type I metric with $h(r)=(1-\frac{(2c)^{4}}{r^{4}}%
)^{1/2}$. In the other extreme case where $k=1$, the larger two $a$'s
coincide and we obtain the Eguchi-Hanson type II metric%
\begin{equation}
ds_{EHII}^{2}=\frac{dr^{2}}{h^{2}(r)}+\frac{r^{2}}{4}h^{2}(r)\sigma _{1}^{2}+%
\frac{r^{2}}{4}(\sigma _{2}^{2}+\sigma _{3}^{2}),  \label{EHII2}
\end{equation}%
which is of the same form of well-known Eguchi-Hanson metric 
\begin{equation}
ds_{EH}^{2} =\frac{r^{2}}{4g(r)}\left[ d\psi +\cos (\theta )d\phi \right]
^{2}+g(r)dr^{2}+\frac{r^{2}}{4}\left( d\theta ^{2}+\sin ^{2}(\theta )d\phi
^{2}\right),
\end{equation}%
by making the substitution $2c=a$ and $h(r)=\frac{1}{\sqrt{g(r)}}$ in (\ref{EHII2}). 
We note that only for special values of $k=0$ and $k=1,$ the metric (\ref{BIX}%
) admits a tri-holomorphic $U(1)$ isometry; hence could be put into
Gibbons-Hawking form. In both special cases of $k=0$ and $k=1$, the
five-dimensional supergravity solutions can be constructed simply by four
harmonic functions on the base space. The case with $k=1$ was considered
explicitly in \cite{To1}, where the authors constructed five-dimensional
supersymmetric black ring solutions on the hyper-K\"{a}hler Eguchi-Hanson
type II base space by. Their solutions have the same two angular momentum
components and the asymptotic structure on time slices is locally Euclidean. 
The circle-direction of the black ring is along the
equator on a two-sphere bolt on the base space. The case with $k=0$ gives a
separable five-dimensional metric for Eguchi-Hanson type I manifold with a time
direction.

By increasing the parameter $k$ as $0<k<$ $1,$ we obtain triaxial Bianchi
type IX metrics with a generic non-tri-holomorphic $U(1)$ isometry. In the
next section, we solve the five-dimensional supergravity equations and find
the solutions.

\bigskip

\bigskip

\section{Supergravity Solutions Over Bianchi Type IX Base Space}

\label{sec:sc}{\Large \ \ \ \ \ }

From equation (\ref{dg}), we can write $G^+=\lambda d \Gamma$ where 
$\Gamma$ is a one-form and $\lambda$ is a 
constant. We take the following ansatz for one-forms $\Gamma$ and 
$\omega$ \cite{Ga1}
\begin{eqnarray}
\Gamma &=& p(r) \sigma _1, \label{Ga} \\
\omega &=& \psi(r) \sigma _1 \label{psi}, 
\end{eqnarray}
where $p$ and $\psi$ are two functions of $r$ (or $\zeta$ through equation (\ref{rzeta})). We find then
\begin{equation}
G^+=2\lambda\big (\frac{2}{r^2}p(r){\bold e}^{(2)}\wedge {\bold e} ^{(3)}-\frac{p'(r)}
{r}
{\bold e}^{(r)}\wedge {\bold e}^{(1)}\big ),\label{Gp}
\end{equation}
where ${\bold e}^{(1)}=\frac{r}{2}\frac{\sqrt[4]{F(r)}}{\sqrt{1-\frac{a_1^4}{r^4}}}\sigma_1$, 
${\bold e}^{(2)}=\frac{r}{2}\frac{\sqrt[4]{F(r)}}{\sqrt{1-\frac{a_2^4}{r^4}}}\sigma_1$,
${\bold e}^{(3)}=\frac{r}{2}\frac{\sqrt[4]{F(r)}}{\sqrt{1-\frac{a_3^4}{r^4}}}\sigma_1$
and ${\bold e}^{(r)}=-\frac{dr}{\sqrt[4]{F(r)}}$ are 
vierbeins for the metric (\ref{BIXG}).
Since $G^+$ is self-dual, we find
\begin{equation}
p(r)=\frac{p_0}{r^2},\label{pp}
\end{equation}
hence we find a simple
analytic forms for one-form $\Gamma$ and self-dual two-form $G^+$
\begin{eqnarray}
\Gamma&=&\frac{p_0}{r^2}
\sigma_1,
\label{pexplicit}\\
G^+&=&\frac{4\lambda p_0}{r^2}
\big ({\bold e}^{(r)}\wedge {\bold e} ^{(1)}+{\bold e}^{(2)}\wedge {\bold e}^{(3)}\big ).\label{exGp}
\end{eqnarray}

%


The Laplace operator on base space (\ref{BIX}), simply is given by 
\begin{equation}
\bigtriangledown ^2=\frac{1}{r^3}\partial _r \{ r^3 \sqrt{F(r)} \partial _r \},
\label{lap}
\end{equation}
so from equation (\ref{eqforF}), we find 
\begin{equation}
\frac{1}{f(r)}=-n\int \frac{dr}{r^7\sqrt{F(r)}} +f_1\int\frac{dr}{r^3\sqrt{F(r)}}+f_2,
\label{5Df}
\end{equation}
where $n=\frac{32}{9}\lambda^2 p_0^2$ and $f_1,f_2$ are two constants.
Although we can't express the metric function $f(r)$ in
a closed analytic form (since the integrals can't be expressed 
in terms of known functions), but we can find all necessary
information about five-dimensional supergravity solutions by
numerically integration.
To determine the one-form $\omega$, we use equation (\ref{Gplus}) together 
with (\ref{Gp}) and we find
the first order differential equation for $\psi(r)$ as
\begin{equation}
\psi '(r)-\frac{2\psi(r)}{r}=-\frac{4\lambda p_0}{r^3f(r)}.
\label{psieq1}
\end{equation}
To solve this differential equation, we multiply it by $1/r$ and then
we find the solution as
\begin{equation}
\psi (r)=-4\lambda p_0 r^2 \int \frac{dr}{r^5 f(r)}+\psi _0 r^2,
\label{psieq2}
\end{equation}
where $\psi _0$ is a constant.
We consider now three different cases corresponding 
to different values for the  parameter $k$, i.e. $k=1$, $k=1$ and $0<k<1$. In the first two cases, the reduced triaxial bianchi type IX space get an enhanced tri-holomorphic symmetry, while in the last case there is no tri-holomorphic symmetry.

First, we consider the special case $k=1$ which triaxial Bianchi type IX
space (\ref{BIX}) reduces to the Eguchi-Hanson type II metric (\ref{EHII2}). As a 
result of reduction, the
non-tri-holomorphic isometry breaks and the reduced space admits an enhanced
tri-holomorphic Killing vector field and it is asymptotically locally Euclidean 
with a self-dual curvature.
The metric has a single removable bolt singularity if $\psi $ is restricted
to the interval $(0,2\pi )$ and the topology of the manifold is $S^{3}/Z_{2}$
asymptotically, hence the manifold is asymptotically locally Euclidean.
We note that near bolt singularity ($r=2c(1+\epsilon ^{2})$, where $\epsilon <<1$),
where the Killing vector
field $\frac{\partial}{\partial \psi}$ vanishes,
the metric reduces to 
\begin{eqnarray}
ds_{r=2c(1+\epsilon ^{2})}^{2} &=&4c^{2}\{\epsilon ^{2}\left[ d\psi +\cos
(\theta )d\phi \right] ^{2}+d\epsilon ^{2}\}+c^{2}\left( d\theta
^{2}+\sin ^{2}(\theta )d\phi ^{2}\right)   \label{dseh4reqa} \\
&\approx &z^{2}d\psi ^{2}+dz^{2}+c^2\left( d\theta ^{2}+\sin
^{2}(\theta )d\phi ^{2}\right).
\end{eqnarray}%
So the space has the topology of $\mathbb{R}^{2}\otimes S^{2}$\ with the radial length equal to $
\sqrt{z^{2}+c^2}$.
If we change the coordinates to
\begin{eqnarray}
R&=&\frac{1}{2c}\sqrt{r^4-16c^4\sin ^2 \theta},\label{C1}\\
\Theta&=&\tan^{-1}\big (\frac{\sqrt{r^4-16c^4}}{r^2}\tan\theta\big ),\label{C2}\\
\Phi&=&\psi,\label{C3}\\
\Psi&=&2\phi,\label{C4}
\end{eqnarray}
where $2c \leq R < \infty , \, 0 \leq \Theta \leq \pi, \, 0 \leq \Phi \leq 2\pi, 
\, 0 \leq \Psi \leq 4\pi$, then the Eguchi-Hanson type II metric transforms into the Gibbons-Hawking two-center form
\begin{equation}
ds_{EHII}^2=H(R,\theta)\big (dR^2+R^2(d\Theta ^2+\sin^2\Theta d\Phi^2)\big )
+\frac{1}{H(R,\Theta )}(\frac{c}{4}d\Psi +Y(R,\theta) d\Phi )^2,
\end{equation}
where
$H(R,\Theta)=\frac{c}{4}\{ \frac{1}{R-R_1}+ \frac{1}{R-R_2} \}
$
and 
$Y(R,\theta)=\frac{c}{4} \big ( \frac{R\cos\theta -2c}{\sqrt{R^2+4c^2-4Rc\cos\Theta}}+
\frac{R\cos\theta +2c}{\sqrt{R^2+4c^2+4Rc\cos\Theta}} \big ).
$
Here $R_1=(0,0,2c)$ and $R_2=(0,0,-2c)$ are Euclidean position vectors of two
nut singularities.

In this special case, the larger two $a$'s in metric function (\ref{FF}) coincide. As a result, we can perform
the integrals in (\ref{5Df}) analytically and we find the 5-dimensional supergravity metric function 
\begin{equation}
1/f=\frac{\mu ^2}{9c^4r^2}-(\frac{\mu^2}{72c^6}-\frac{f_1}{16c^2})\big ( \ln(r^2+4c^2)-
\ln(r^2-4c^2) \big )+f_2,\label{fEHII}
\end{equation}
and 
\begin{equation}
\psi (r) = \frac{2\mu ^3}{27c^4r^4}+\frac{\mu f_2}{r^2}+\frac{\mu f_1}{32c^4}-
\frac{\mu ^3}{144 c^8}+\psi _0 r^2-\frac{\mu(9 f_1 c^4-2 \mu ^2)}{2304r^2c^{10}}
(r^4-16c^4)\ln ( \frac{r^2+4c^2}{r^2-4c^2} ),
\end{equation}
where $\mu=\lambda p_0$.
These results are exactly in agreement\footnote{We should note that in \cite{G1}, the authors used the most negative 
signature for the five dimensional metric.} with the results of \cite{G1}. 

To have a regular solution at $r=2c$, we should set $f_1=\frac{2\mu ^2}{9c^4}$ and then
to have a positive definite metric function, we should choose 
$f_2 \geq -\frac{\mu ^2}{36 c^6}$. Moreover, to eliminate the time-like curves at 
$r\rightarrow \infty$, we should set $\psi _0=0$.
However, there are always closed time-like curves near $r=2c$.


Second, we consider the special case $k=0$ which triaxial Bianchi type IX
space (\ref{BIX}) reduces to the Eguchi-Hanson type I metric (\ref{EHI}). As a 
result of reduction, the
non-tri-holomorphic isometry again breaks and the reduced space admits an enhanced
tri-holomorphic Killing vector field.
In this case, to get a real five-dimensional metric function $f$ and one-form $\psi$, 
we should choose $\mu=\psi _0=0$.
So, the five dimensional metric function becomes a constant $f_2$, and the 5-d metric is just 
$-f_2^2dt^2+\frac{1}{f_2}ds_{EHI}^2$.


Third, we consider triaxial Bianchi type IX space with a generic $0<k<1$. 
Although we can't integrate the integrals in equation (\ref{5Df}) to find an analytic form for the
metric function, but we can integrate numerically. The behavior of metric function for different values of parameters are plotted in fig 1 and 2.
In fig 1, we choose $f_1$ to be zero while in fig 2, $f_1\neq 0$. In both cases, for large $r$ the metric function approaches a constant. We should note that changing the value of $k$ doesn't change the decaying behavior of function $f$ for $f_1=0$, as well as no change for behavior of function $f$ for $f_1 \neq 0$. The change of parameter $k$ only shifts the value of metric function $f$.
\begin{figure}[tbp]
\centering           
\begin{minipage}[c]{.3\textwidth}
        \centering
        \includegraphics[width=\textwidth]{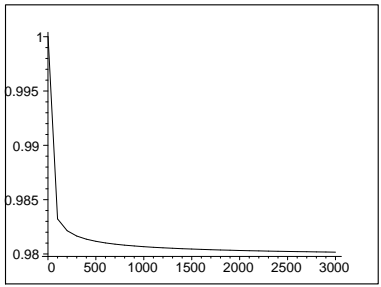}
    \end{minipage}
\caption{
The metric function $f$ as a function of $\frac{1}{r-2c}$.
we set $f_1$=0 and $f_2$=1.
}
\label{fig1}
\end{figure}
\begin{figure}[tbp]
\centering           
\begin{minipage}[c]{.3\textwidth}
        \centering
        \includegraphics[width=\textwidth]{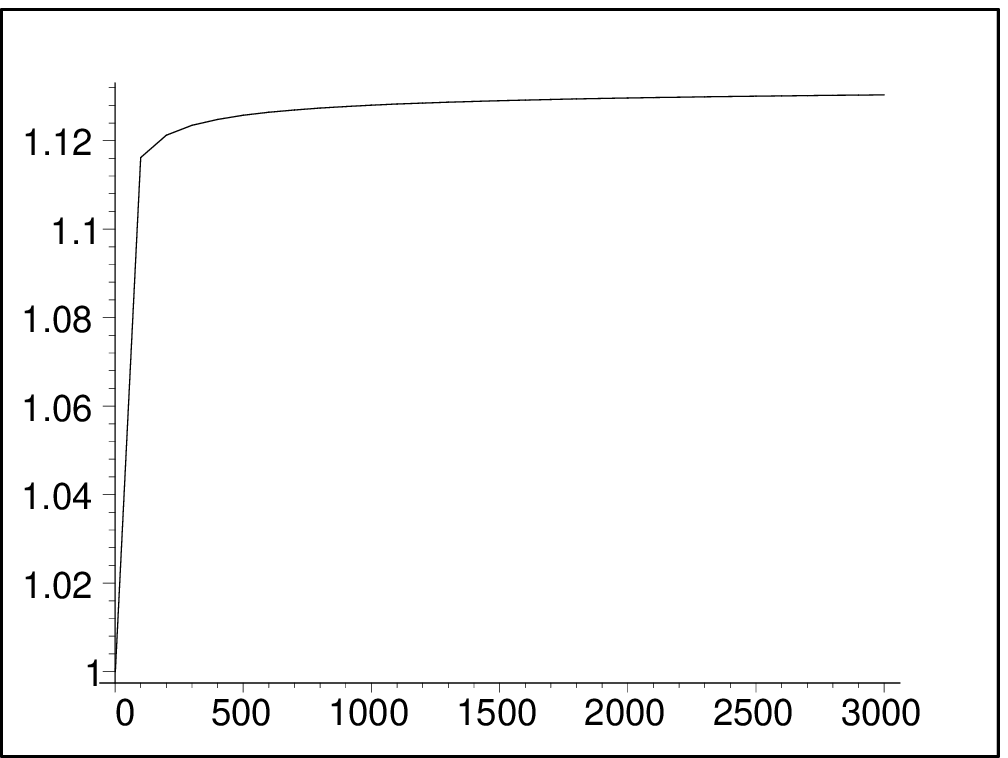}
    \end{minipage}
\caption{
The metric function $f$ as a function of $\frac{1}{r-2c}$.
we set $f_1$=$f_2$=1.
}
\label{fig2}
\end{figure}

In fact, in the large $r$ limit, the metric function behaves as  
\begin{equation}
f(r) \sim \frac{1}{f_2}(1+\frac{f_1}{2f_2r^2})+ O(\frac{1}{r^4}),
\end{equation}
hence, asymptotically, the five dimensional metric is
\begin{equation}
ds^2\rightarrow -\frac{1}{f_2^2}(dt+\frac{\lambda p_0f_2}{r^2}\sigma_1)^2+f_2ds_{tri.\text{ }B.\text{ }IX}^2.
\end{equation}

By looking at the coefficient of $g_{\psi\psi}$, we see $\partial /\partial \psi$ becomes time-like if
$4f^3(r)\psi^2(r) > r^2/4\sqrt{F(r)}$, hence we get closed-timelike curves. At large
-r limit, $\psi(r) \rightarrow \frac{\mu f_2}{r^2}$ and $f(r) \rightarrow \frac{1}{f_2}$
, hence we conclude there are no closed-timelike curves if $r$ is quite large. 
On the other extreme limit, close to $r=2c$; the metric function $F(r)$ has
Taylor expansion 
\begin{equation}
\sqrt{F}\simeq F_1\sqrt{\epsilon}+F_2\epsilon^{3/2}+O(\epsilon^{5/2}),
\end{equation}
where $F_1=\sqrt{2\frac{1-k^4}{c}}$,
$F_2=\frac{F_1}{8c}\frac{5-13k^4}{k^4-1}$ and $\epsilon=r-2c$.
So from equation (\ref{5Df}), we find the
following behavior for the five-dimensional metric function $f(r)$ near $r=2c$ 
\begin{equation}
f(r) \simeq \frac{1}{f_2}+\frac{n-16c^4f_1}{64f_2^2c^7F_1}\sqrt{\epsilon}+O(\epsilon).
\end{equation}
This result for the metric function $f(r)$ with a generic $0< k <1$ is remarkable since it provides a smooth transition from the metric function of Eguchi-Hanson type I based solution with $k=0$ to the metric function of Eguchi-Hanson type II based solution with $k=1$. 

In the limit of $r\simeq 2c$, the field $\psi(r)$ behaves as 
\begin{equation}
\psi(r) \simeq (\frac{\lambda p_0 f_2}{4c^2}+4\psi _0 c^2)+(-\frac{\lambda p_0 f_2}{4c^3}+4\psi _0 c)\epsilon +O(\epsilon ^{3/2}).
\end{equation}

To summarize, we have found the five-dimensional supergravity equations of motion. These equations lead to differential equations for the fields $f(r)$ and $p(r)$ and $\psi(r)$. We have solved the equations for $r\sim 2c$ and at infinity and also found the numerical solutions for the other values of $r$.





\section{Conclusions}

We have constructed a new class of solutions to five dimensional supergravity, based on Bianchi type IX Einstein-hyperk\"{a}hler space. The Bianchi type IX Einstein-hyperk\"{a}hler space doesn't have any tri-holomorphic $U(1)$ isometries, hence the
solutions could be used to study the physical processes that don't respect
any tri-holomorphic abelian symmetries. We find the solutions to the equations of motion near $r=2c$
and at infinity. Moreover, by numerical integration, we explicitly find the general behavior of the solutions. Our solutions based on triaxial Bianchi 
type IX space provides a smooth transition from solution based on Eguchi-Hanson type I space to corresponding solution based on Eguchi-Hanson type II space.
One feature of new solution is that in various limits of parameters,
it reduces to many of previously constructed
five-dimensional supergravity solutions based on both hyper-K\"{a}hler base spaces
that can be put into a Gibbons-Hawking form and hyper-K\"{a}hler base spaces that
can't be put into a Gibbons-Hawking form.

We conclude with a few comments about possible directions for future work. 
We have shown that our new solution (based on tri axial Bianchi type IX space) is asymptotically free of any closed time-like curves. An interesting application of our solutions is to seek for their holographic dual theories. In the case of existence of boundary holographic dual theories to the solutions (or their generalization with cosmological constant), then they would be free of any irregularities associated with the closed time-like curves in the bulk of space-time.  
Moreover, in deriving our solutions, we have considered the simplest dependence of the five dimensional metric function on the coordinates (i.e. dependence only to the radial coordinate).
The reason for taking the simplest dependence is due to the non-homogeneity of differential equation (\ref{eqforF}) for the metric function, otherwise finding the solutions could be very hard or even impossible. If we want to find some other solutions (with metric function depending on two or more coordinates), then we should consider the homogenous differential equation ((\ref{eqforF}) with   $G^+$ equal to zero). It's quite possible that in solutions to the homogenous differential equation, the point $r=2c$ can be converted to a regular hypersurface(s) in five dimensional space-time and we obtain black hole solutions.
The other open issue is study of the thermodynamics of constructed solutions in
this paper.

\bigskip
\bigskip
\noindent{\Large {\bf Acknowledgments}}
\newline

This work was supported by the Natural Sciences and Engineering Research
Council of Canada (NSERC).

\bigskip

\noindent{\Large {\bf Appendix}}
\newline

In this appendix, we collect some formulas for Maurer-Cartan one-forms, Jacobi Theta and elliptic functions which are crucial for our discussion in section 3. 

First of all, the Maurer-Cartan one-forms $\sigma _{i\text{ }}$ are given by
\begin{eqnarray}
\sigma _{1}&=&d\psi +\cos \theta d\phi, \label{s1}\\ 
\sigma _{2}&=&-\sin \psi d\theta +\cos \psi \sin \theta d\phi,\label{S2} \\ 
\sigma _{3}&=&\cos \psi d\theta +\sin \psi \sin \theta d\phi, \label{S3}%
\end{eqnarray}%
with the property 
\begin{equation}
d\sigma _{i}=\frac{1}{2}\varepsilon _{ijk}\sigma _{j}\wedge \sigma _{k}.
\label{dsigma}
\end{equation}%
We note that the metric on the $\mathbb{R}^{4}$ (with a radial coordinate $R$
and Euler angles ($\theta ,\phi ,\psi $) on an $S^{3}$) could be written in
terms of Maurer-Cartan one-forms via%
\begin{equation}
ds^{2}=dR^{2}+\frac{R^{2}}{4}(\sigma _{1}^{2}+\sigma _{2}^{2}+\sigma
_{3}^{2}),  \label{s3METRIC}
\end{equation}%
with $\sigma _{1}^{2}+\sigma _{2}^{2}$\ is the standard metric of $S^{2}$\
with unit radius; $4(\sigma _{1}^{2}+\sigma _{2}^{2}+\sigma _{3}^{2})$\
gives the same for $S^{3}.$ 

The Jacobi Theta functions $\vartheta $ are given by
\begin{equation}
\vartheta _{i}(\tau )=\vartheta _{i}(0\left| \tau \right. ),  \label{th}
\end{equation}%
where we have used Jacobi-Erderlyi notation%
\begin{eqnarray}
\vartheta _{1}(\nu \left| \tau \right. ) &=&\vartheta \left[ _{1}^{1}\right]
(\nu \left| \tau \right. ),  \label{thetas} \\
\vartheta _{2}(\nu \left| \tau \right. ) &=&\vartheta \left[ _{0}^{1}\right]
(\nu \left| \tau \right. ),  \label{th1} \\
\vartheta _{3}(\nu \left| \tau \right. ) &=&\vartheta \left[ _{0}^{0}\right]
(\nu \left| \tau \right. ),  \label{th2} \\
\vartheta _{4}(\nu \left| \tau \right. ) &=&\vartheta \left[ _{1}^{0}\right]
(\nu \left| \tau \right. ).  \label{th3}
\end{eqnarray}
The Jacobi functions with characteristics 
$\vartheta \left[ _{b}^{a}\right] (\nu \left| \tau \right. )$ are defined by the following series
\begin{equation}
\vartheta \left[ _{b}^{a}\right] (\nu \left| \tau \right. )=\sum_{n\in
Z}e^{i\pi (n-\frac{a}{2})\{\tau (n-\frac{a}{2})+2(\nu -\frac{b}{2})\}},
\label{thy}
\end{equation}%
where $a$ and $b$ are two real numbers.
%

The standard Jacobi elliptic $SN$, $CN$ and $DN$ functions $\mathfrak{sn}(z,k)$, $\mathfrak{cn}(z,k)$ and $\mathfrak{dn}(z,k)$, are related respectively,
to $\mathfrak{am}(z,k)$; the Jacobi elliptic $AM$ function, by%
\begin{eqnarray}
\mathfrak{sn}(z,k)&=&\sin (\mathfrak{am}(z,k)),  \label{SN}\\
\mathfrak{cn}(z,k)&=&\cos (\mathfrak{am}(z,k)),  \label{CN}\\
\mathfrak{dn}(z,k)&=&\sqrt{1-k^2\mathfrak{sn}^2(z,k)}.  \label{DN}
\end{eqnarray}%
The Jacobi elliptic $AM$ function is the inverse of the trigonometric form
of the elliptic integral of the first kind; which means%
\begin{equation}
\mathfrak{am}(\mathfrak{f}(\sin \phi ,k),k)=\phi,   \label{AM}
\end{equation}%
where $\mathfrak{f}(\varphi ,k)$; the elliptic integral of the first kind, is
given by 
\begin{equation}
\mathfrak{f}(\varphi ,k)=\int_{0}^{\sin ^{-1}(\varphi )}\frac{d\theta }{%
\sqrt{1-k^{2}\sin ^{2}\theta }}.  \label{F}
\end{equation}
The Eguchi-hanson type I matric also can be written as \cite{EH} 
\begin{equation}
ds_{EHI}^{2}=\widetilde{f}^{2}(r)dr^{2}+\frac{r^{2}}{4}\widetilde{g}%
^{2}(r)\{d\theta ^{2}+\sin ^{2}\theta d\phi ^{2}\}+\frac{r^{2}}{4}(d\psi
+\cos \theta d\phi )^{2},  \label{EHIs}
\end{equation}%
where%
\begin{equation}
\begin{array}{c}
\widetilde{f}(r)=\frac{1}{2}(1+\frac{1}{\sqrt{1-\frac{a^{4}}{r^{4}}}}), \\ 
\widetilde{g}(r)=\sqrt{\frac{1}{2}(1+\sqrt{1-\frac{a^{4}}{r^{4}}})}.%
\end{array}
\label{EHIsmff}
\end{equation}%

\end{document}